# DFT study of pressure induced phase transitions in LiYF$_4$

B.Minisini, L. El Hadj, M. Lontsi Fomena, N. Van Garderen and F. Tsobnang.

Institut Supérieur des Matériaux et Mécaniques Avancés du Mans, 44 Av. Bartholdi, 72000 Le Mans, France

Email : bminisini@ismans.univ-lemans.fr

**Abstract :**

An investigation of the pressure induced phase transition from the scheelite phase (I4$_1$/a, Z=4) to the fergusonite-like phase (I2/a, Z=4)/LaTaO(P2$_1$/c, Z=4) of LiYF$_4$ is presented. Employing density functional theory (DFT) within the generalized gradient approximation, the internal degrees of freedom were relaxed for a pressure range of 0 GPa to 20 Gpa. The influence of pressure on the lattice vibration spectrum of the scheelite phase (I4$_1$/a, Z=4) was evaluated using the direct approach, i.e. using force constants calculated from atomic displacements. The transition volume is in good agreement with experiment, while the transition pressure is overestimated of 6 GPa. At 20 GPa, a P2$_1$/c structure with apentacoordinated lithium cation is found to be the most stable phase. This structure is compatible with a transition driven by a B$_g$ zone-center soft optic mode linked to a soft-acoustic mode along the [11-1] direction as observed for the proper ferroelastic transition of BiVO$_4$.

## 1. Introduction

Due to its importance as a host materials for laser applications, undoped or doped LiYF$_4$ was extensively studied during the last years[1]-[3]. Concerning the phase transition is now commonly accepted that from 30 K to 1000 K scheelite phase (I4$_1$/a, Z=4) is stable[4]. Most of the recent studies were performed at ambient pressure conditions, so the transition pressures as well as the high-pressure phases remain to be investigated in more detail.

Blanchfield et al.[5], did hydrostatic pressure studies from 0 GPa to 16 GPa on second-order elastic constantsand thereby showed the first acoustic mode softening. However no information was obtained concerning phase transitions. More recently, Sarantopoulou et al.[4] noticed a discontinuity in the Raman active modes as a function of the pressure at 7 GPa. Changes of Li-F distances rather than $Y^{3+}$-$LiF_4^{-3}$ were assumed to be the cause of this anomaly, the structure remaining tetragonal. No other changes were observed up to a pressure of 16 GPa. From X-ray powder diffraction measurement performed on $LiYF_4$ and on $CaWO_4$ Errandonea et al.[6] proposed the existence of a reversible phase transition of two scheelite polytypes induced by $LiF_4$ polyhedral tilting at 6 GPa. Moreover, Grzechnik et al.[7] identified two pressure induced phase transitions. The first one occurs at 10 GPa and was supposed to lead to fergusonite (I2/a, Z=4) type structure. The second one was detected around 17 GPa. It was impossible for the authors to determine reliable data for cell parameters but they drew up a list of possible crystallographic structures.

Based on these experimental data, some theoretical studies were undertook. Sen et al.[8][9] observed two phase transitions using a empirical rigid ion model (RIM). The first one around 5 GPa was supposed to be a second order phase transition without volume drop. Moreover the crystallographic structure was observed to depend on temperature. Below 400 K the LaTaO($P2_1/c$, Z=4) type phase was found to be the most stable whereas above this temperature the Fergusonitelike (I2/a, Z=4) structure was more stable. The second transition around 16 GPa was found to be a first order transition phase with a volume drop of 6% leading to a LaTaO ($P2_1/c$, Z=4) structure. However, employing density functional theory (DFT), Li et al.[10] found a first transition at 0 K and 9.3 GPa leading to fergusonite (I2/a, Z=4) type and a second one at 17.6 GPa leading to wolframite (P2/c, Z=2) type.

In this work we present a DFT analysis of structural properties of $LiYF_4$ scheelite ($I4_1/a$, Z=4), fergusonite (I2/a, Z=4), wolframite (P2/c, Z=2) type and $LaTaO_4$ ($P2_1/c$, Z=4) types under pressure at 0K. The structures were optimized under pressure ranging from 0 GPa to 20 GPa without imposing any symmetry restrictions. Structural parameters and x-ray powder diffraction diagrams of optimized cells were analyzed. Enthalpies of formation were calculated to evaluate the most stable phase under pressure. In addition vibrational properties of the optimized structures from initial scheelite ($I4_1/a$, Z=4) type were calculated under pressure up to 20 GPa.

## 2. Simulation

All calculations were carried out using density functional theory (DFT) [11][12] as implemented in the Vienna Ab-Initio Simulation Package (VASP) [13], part of the MedeA modelling interface[14].

*2.1 Structure optimisation*

Experimental structural parameters for scheelite ($I4_1/a$,z=4)[15] and fergusonite ($I2/a$, z=4) [7], were used to build our models. The $P2_1/c$ (Z=4) was built from lattice data obtained numericaaly by Sen et al.[9]. The 1*2*1 supercell of wolframite ($P2/c$, z=2) was constructed based on calculated data by Li et al.[10]. In the literature cited above, the labelling for the tetragonal cell axis is different between the structures. For clarity we have therefore renamed the unit cell axis *a, b, c* of the $I4_1/a$ structures to *c, a* and *b* respectively.

The exchange correlation functional was approximated by the gradient corrected form proposed by Perdew and Wang[16]. The electronic degrees of freedom were described using the projector augmented wave method(PAW)[17][18]. Electronic convergence was set up at $10^{-6}$ eV. Sampling of the Brillouin zone were performed by the Monkhorst-Pack scheme[20]. A convergence test concluded that a k-spacing of 0.5 Å$^{-1}$ was sufficient to obtain a total energy convergence within 0.1 meV/cell compared to 0.2 Å$^{-1}$ at 0 GPa. We found a plane wave cutoff of 700 eV to be sufficient for the convergence of structural parameters. The total energy changes by less than 3 meV/LiYF$_4$ on increasing the cutoff from 700 eV to 750 eV.

Full structural relaxations of the three initial structures defined earlier were performed at 0, 2, 4, 6, 10, 12, 14, 16, 18, 20 GPa except for the wolframite supercell where a single calculation was performed at 20 GPa. All atoms were considered to be symmetrically inequivalent in order to allow for potential symmetry changes due to pressure. Calculations were considered converged when residual forces were less than 2 meV/Å. At 0 GPa, the residual bulk stresses were smaller than 25 MPa for all structure. To define the optimized structures we define a nomenclature as followed; $S_P^{sym}$. The subscript *sym* stands for the symmetry of the structure before optimisation, the *P* subscript stands for the pressure imposed during the optimisation.

## 2.2 Lattice dynamic calculation

The lattice vibrational properties were calculated within the harmonic approximation, using the PHONON code[21], based on the harmonic approximation. Using PHONON the force constant matrix was calculated via atomic displacements with an interaction range of 7 Å. The asymmetric atoms were displaced by +/- 0.03 Å leading to 14 new structures. The dynamical matrix was obtained from the forces calculated via the Hellmann-Feynman theorem. The selected k-point spacing led to 6 symmetry independent 6 symmetry-unique k-points a planewave cutoff of 550 eV was used to describe the electronic valence states. The error on the force can perturb the translation-rotational invariance condition. Consequently, this condition has to be enforced. A strength of enforcement of the translational invariance condition was fixed at 0.1 during the derivation of all force constants. The longitudinal optical (LO) and transversal optical mode(TO) splitting was not investigated in this work. Consequently, only TO modes at the **Γ** point were obtained.

## 2.3 Mechanical properties calculation

Elastic constants have been calculated for $S_P^{I4_1/a}$ from the stress evaluated on strained cells. The strain level was set up at 1% and we used the same calculation parameters as in section 2.1. For crystal belonging to TII Laue group, the elastic stiffness constant matrix referred to the crystallographic axis (X, Y, Z) reference frame is.

$$C_{ij} = \begin{pmatrix} c_{11} & c_{12} & c_{13} & 0 & 0 & c_{16} \\ c_{12} & c_{22} & c_{13} & 0 & 0 & -c_{16} \\ c_{13} & c_{13} & c_{33} & 0 & 0 & 0 \\ 0 & 0 & 0 & c_{44} & 0 & 0 \\ 0 & 0 & 0 & 0 & c_{44} & 0 \\ c_{16} & -c_{16} & 0 & 0 & 0 & c_{66} \end{pmatrix}$$

The sign of the elastic coefficient $C_{16}$ depends on the choice of the +Z axis or +Y axis respectively before and after relabelling. To evaluate the sense of the axis we used the standard convention employed for the sheelite structures[22]. Complete discussion about this topic is available elsewhere[23].

## 3. Results

*3.1 Structural properties*

The pressure dependence of the lattice parameters are presented in Figure 1. From these results it is worth noting that during the lattice optmization the three different initial structures converged to one and the same cell parameters for pressures below 10 GPa. The equilibrium cell parameters at 0 GPa are $a=c=5.23$ Å, $b=10.83$ Å and $\alpha=\beta=\gamma=90°$, giving a cell volume of 296 Å$^3$. This volume is 3.7% higher than the experimental one, confirming the overcorrection of the GGA on local density approximation as already discussed by Li et al.[10] for this structure. The experimental volume of 286 Å$^3$ is obtained numerically for an applied pressure of between 4 GPa and 6 GPa.

At 10 GPa, the *a* and *c* parameters take two distinct values for $S_{P=10}^{I2/a}$ and $S_{P=10}^{P2_1/c}$ whereas these parameters remain identical for $S_{P=10}^{I4_1/a}$. Moreover, the monoclinic angle *β* jumps to 94° for $S_{P>10}^{P2_1/c}$ to converge to 97° at 20 GPa. The evolution is more progressive for $S_{P>10}^{I2/a}$ even if beyond 15 GPa *β* exceeds 97°. The difference in volume of the three optimized structures below 8 GPa does not exceed 0.1%. At 16 GPa, the volume of $S_{P=16}^{I4_1/a}$ is 1.78% higher than that of $S_{P=16}^{I2/a}$ and 1.96% % higher than that of $S_{P=16}^{P2_1/c}$.

The X-ray pattern obtained from these structures are plotted in Figure 2. Until 10 GPa, the spectra of the optimized structures from initial $S_{P<10}^{I4_1/a}$, $S_{P<10}^{I2/a}$, $S_{P<10}^{P2_1/c}$ are identical. As result of the structural parameters changes observed from 10 GPa upwards, a peak splitting at 2? = 5.11° appears for $S_{P>10}^{P2_1/c}$ and $S_{P>10}^{I2/a}$. The splitting remains over the whole pressure range and the congruence of the two patterns is striking. At 20 GPa, the difference between the second peak and the third peak is 2.29° in agreement with the measured value[7]. Concerning the wolframite $S_{P=20}^{P2/c}$ at 20 GPa, the largest difference between the first and the second peak is 1.40°. Such a difference being associated with the general shape of the pattern seems incompatible with the experimental patterns[7]. Moreover, from recent investigation[26] it would seem that the transformation in wolframite is opposed to the natural sequence of transformation for this group of materials. For these reasons we excluded this structure from further considerations.

The interatomic distances as a function of the pressure for $S_P^{P2_1/c}$ are given in the Figure 3. At 0 GPa, the Li-F bonds distances in the LiF$_4$ tetrahedra were 1.92 Å. In the YF$_8$ octahedra, the four first neighbours are situated at 2.26 Å and the distance to the four second neighbours is 2.32 Å. At 20 GPa, the Li-F distances are lie between 1.79-1.84 Å. Further, one of the second neighbours initially situated at 2.93 Å is now at 1.95 Å. Consequently, Li can be assumed as penta-coordinated, the tetrahedral being transformed into a distorted trigonal bipyramidal as shown in Figure 4. Concerning Y-F distances at 20 GPa, the distances of the 8 nearest neighbors are shorter compared to low pressure. So, following the convention in the litterature.[6], the cation coordination of LiYF$_4$ is (8-5). This result confirms the hypothesis of the existense of an intermediate structure between (8-4) and (8-6) suggested by Errandonea et al.[6]. However this structure is different from that calculated by Sen and Chaplot[8] leading to Li and Y octahedrally and tenfold coordinated to the fluorine atoms respectively.

The evolution of the angle $F$ defined by Blanchfield et al.[23] as a function of the pressure is presented in Figure 5 for all three systems studied. A value of 30,3° is obtained for $S_{P=0}^{I2/a}$ and $S_{P=0}^{P2_1/c}$ at ambiant pressure which is in agreement with the experimental value. However, the angle evaluated for $S_{P=0}^{I4_1/a}$ is 10% higher than the other structures. Concerning $S_{P}^{I4_1/a}$, the evolution of $F$ is progressive with an increase of 0.2° per 2 GPa all over the explored pressure range. For $S_{P}^{I2/a}$ we notice a discontinuity at 14 GPa and 16 GPa with an increase going up to 0.6°. Concerning $S_{P}^{P2_1/c}$, the trend is different. Between 0 and 12 GPa, the angle becomes larger with the most important increase at 10 GPa. Then from 14 GPa upwards the angle decreases up to 30.7° at 20 GPa.

*3.2 Enthalpies*

Figure 6 shows enthalpies H for the systems $S_{P}^{I2/a}$ and $S_{P}^{P2_1/c}$ relative to $S_{P}^{I4_1/a}$. Below 14 GPa, $S_{P<14}^{I4_1/a}$ turned out to be the most stable at 0K. From 0 GPa to 10 GPa, the difference in enthalpy is completely due to the internal energy since the volumes are the same for all structures. Between 10 GPa and 14 GPa, the enthalpy of $S_{10\leq P\leq14}^{I2/a}$ remains stable whereas the enthalpy of $S_{10\leq P\leq14}^{P2_1/c}$ approaches that of $S_{10\leq P\leq14}^{I4_1/a}$. From 16 GPa upwards, the initial $S_{P\geq16}^{P2_1/c}$ structure becomes the most stable one. Over the whole pressure range explored in this study $S_{P}^{I2/a}$ is less favourable than the other phases.

*3.3 Mechanical properties*

Using Voigt's formalism for both numerical and experimental data the mechanical properties were calculated from the elastic constants at ambient pressure. From the result presented in Table 1, we can see that the agreement between the modules is very good. The bulk modulus (B) is important for the description of a crystal behaviour under pressure. From fitting with an equation of state, B was evaluated at 80 GPa[5][7], 94.8 GPa[10] and 69 GPa[8] from experimental, DFT and empirical data respectively. From the elastic constants we evaluated B at 81 GPa to be in very good agreement with the experimental value. The pressure dependence of the elastic constants for $S_P^{I4_1/a}$ is presented in Figure 7. The elastic constants $C_{16}$ and $C_{66}$ and $C_{44}$ show a small pressure dependence. as experimentally measure on a range from [0-0.16] GPa. At 8 GPa, we can notice a discontinuity in the evolution of the elastic constants as a function of pressure. This discontinuity is striking mainly for $C_{11}$ since after an initial weak increase we can notice a decrease from 8 GPa upwards. Moreover, at this pressure one eigenvalue of the elastic constant matrix is negative, indicating a mechanical instability[24].

*3.4 Lattice dynamic*

To study the mechanism of the pressure induced phase transition, we calculated the phonon dispersion curves of $S_P^{I4_1/a}$ within a range from 0 GPa to 20 GPa. First the pressure dependence of the Raman active modes are presented in Figure 8. Below 250 cm$^{-1}$, the frequency of the Eg mode decreases with increasing pressure. This behaviour was already observed by Wang et al. [25] for pressure up to 17 GPa. The corresponding mode was associated with a rotation of the LiF$_4$ tetrahedra in the *ac plane*[29]. However from our analysis of this mode, the rotation of the tetrahedra is around the *c axis*. In fact the rotation of the LiF$_4$ tetrahedra in the *ac plane* is associated with the Ag mode. This mode and the Bg mode, associated to a translation of lithium along the *b axis,* increase for a pressure up to 8 GPa and decreased for higher pressure.

For the vibrational modes above 250 cm$^{-1}$ at 0 GPa, the agreement with experimental values by Sarantopoulou et al.[4] is correct since except for the second Bg mode the numerical values are less than 10% lower than experimental one. The error for the Bg mode is of 13%. Three doublets were observed experimentally above 250 cm$^1$, namely (Eg+Bg) modes at 326

cm$^{-1}$ and 375 cm$^{-1}$ and the (Ag+Bg) mode at 426 cm$^{-1}$. From our calculations, theses doublets are separated even at 0 GPa. The relative difference between the first couple (Eg+Bg) remains constant. This behaviour deviates from experiment since the split of theses two modes lead to a difference higher than 40 cm$^{-1}$ at 15 GPa. Concerning the second doublet (Eg+Bg), from 2 GPa upwards the relative variation increases. From Sarantopoulou et al.[4] experiments, the split of this doublet is very low and does not exceed 20 cm$^{-1}$ at 15 GPa. But from Wang et al.[25] experimental results, the splitting begins at 5 GPa and the difference between the two modes exceeds 20 cm$^{-1}$ at 15 GPa. The splitting of the last doublet (Ag+Bg) takes place around 8 GPa in agreement with experimental finding[4]. But this different behaviour was not observed by Wang et al.[25]. We calculated the $\partial\omega/\partial P$ slopes in the pressure interval [0-8] and [10-16] GPa, the results are presented in Table 2. At a frequency of 246 cm$^{-1}$, the calculated slopes agree very well with experimental values at low pressure. We can notice a discontinuity of the slope from 8 GPa. Except for the high frequency Bg mode, all the slopes are lower at higher pressure agreement with experiment[4][25].

The phonon dispersion curves for different pressures are presented in Figure 9. At 4 GPa an energy gap appears at high frequency. From 8 GPa upwards, we can observe imaginary frequencies close to the Brillouin zone center along the [11-1] direction. These frequencies correspond to a translation of the whole structure along the *a axis* of the scheelite structure. All the frequencies are negative at 16 GPa. At the N point the LiF$_4$ tetrahedra tend to rotate around the *c axis* and the Li atoms start to translate along this axis. Moreover, for this same pressure additional imaginary frequencies appear close to the Brillouin zone center along the [001] direction and another energy gap appears at high frequency. Consequently, the phase transition should be caused by a dynamical instabilities of transverse acoustic phonon modes.

**IV Discussion.**

The computed enthalpy points to a phase transition at a pressure of around 16 GPa. Below this pressure, the $S_{P\leq16}^{I4_1/a}$ structure is the most stable one. At 16 GPa, the difference in enthalpy is less than 1 kJ/mol per LiYF$_4$ between $S_{P=16}^{I2/a}$, $S_{P=16}^{P2_1/c}$ and $S_{P=16}^{I4_1/a}$. From such a small energy difference it is difficult to come to a conclusion on the phase stability at this pressure. Indeed, even if the electronic energy is very well converged, at high pressure the error due to the Pulay stress induces an error around 1 kJ/mol per LiYF$_4$. Moroever, the effect of the temperature can play a significant role as previously noticed by empirical dynamic molecular calculations[9]. In order to take into account the effect of temperature the Gibbs Free Energy needs to be evaluated for all system under investigation. This is a laborious task for the low symmetry system and therefore this analysis was not included in the current analysis.

From 18 GPa upwards, the difference inenthalpy exceed 1 kJ/mol so $S_{P>18}^{P2_1/c}$ seems the most stable at 0K. The calculated transition pressure is around 6 GPa higher than the experimental value determined around 10.3 GPa[7]. In a previous DFT study[27] a similar overestimation was found to be a consequence of the GGA. However, the transition volume is in good agreement with experimental results. Indeed, the experimental volume at 0 GPa is numerically obtained for a pressure between 4GPa and 6 GPa. Moreover if we compare the experimental structure at 10.3 GPa we have a=c=4.95 Å, b=10.50 Å and V=258.33 Å[7], these parameters corresponds to $S_{P=16}^{I4_1/a}$ at 16 GPa (a=c=4.93 Å, b=10.53 Å and V=256.89 Å). From a full Rietveld refinement of the X ray diffraction pattern of LiYF$_4$ taken at 13.3 GPa, the structure was derived to be I2/a with structural parameters *a*=5.04 Å, *c*=4.78 Å, *b*=10.41 Å, *ß*=95.27 Å and *V*=250.03 Å[7]. The cell parameters evaluated for the $S_{P=18}^{P2_1/c}$ at 18 GPa (*a*=5.07 Å, *c*=4.73 Å, *b*=10.40 Å, *ß*=96.49 Å and *V*=248.23 Å) are very closed to the experimental ones.

From 16 GPa upwards, the frequencies associated at the transverse acoustic phonon mode along the [1 1 –1] direction are completely negatives. The softness of this mode appears around 8 GPa and is clearly very visible above 10 GPa. It is associated with a softness of the Ag, Eg and Bg Raman active modes at lower frequencies. Taking into account the shift of 4-6 GPa, this soft modes could explain the anomalies observed around 7 GPa by Raman[4][25] and luminescence[30] measurements. These three modes induce the rotation of the $LiYF_4$ not only in the *ac plane* but also around the *c* axis. These tilting movements can explain the formation of pentacoordinated lithium observed in $S_{P>16}^{P2_1/c}$ at high pressure.

Consequently, we conclude that the transition is driven by a Bg soft optic modes coupled with a soft acoustic mode. This mechanism was previously observed for the $BiVO_4$ and was associated to a proper ferroelastic phase transition[28]. This could confirm that $LiYF_4$ undergoes a pressure-induced ferroelastic phase transition as previously described[26]. To evaluate the order of the phase transition using the theoretical model from Grzechnik et al.[26], further calculations at higher pressure are required

**V Conclusion**

DFT structure optimization of the $LiYF_4$ structure in the pressure range from 0 GPa to 20 GPa point to one pressure induced phase transition. While our calculation overestimate the transition pressure by 6 GPa, the agreement between calculated and experimental transition volume is very good. Raman and luminescence anomalous measurements around 7 GPa can be explained by the softness low frequency modes Ag, Bg and Eg. Theses three modes are associated with a tilting movement of the $LiF_4$ tetrahedra. We found a soft mode assisted transition similar to the temperature induced proper ferroelastic transition observed for $BiVO_4$. Further calculations are required to evaluate the order of this phase transition and to confirm the temperature dependence of the post scheelite phase.

**Acknowledgment**

The authors whish to thank the Materials Design team for the numerical support. We are grateful to CCI du Mans et de la Sarthe for providing calculation infrastructure. Special thanks to A. Mavromaras for discussion.

**Table 1 : Elastic stiffness constants and mechanical moduli of scheelite LiYF$_4$ at 0 GPA.**

|                | Calc.   | Exp.[23] |
|----------------|---------|----------|
|                | Elastic constants ||
| $C_{11}$       | 114±2   | 121      |
| $C_{12}$       | 53±2    | 60.9     |
| $C_{13}$       | 61±1    | 52.6     |
| $C_{16}$       | -11±2   | -7.7     |
| $C_{33}$       | 152±2   | 156      |
| $C_{44}$       | 37±3    | 40.6     |
| $C_{66}$       | 22±3    | 17.7     |
|                | Modulus ||
| *B* Bulk (GPa) | 81      | 81       |
| *G* Shear (GPa)| 33      | 35       |
| *E* Young (GPa)| 87      | 92       |
| *n* Poisson    | 0.32    | 0.31     |

$G=1/15(c_{11}+c_{22}+c_{33}-c_{12}-c_{13}-c_{23})+1/5(c_{44}+c_{55}+c_{66})$
$B=1/9(c_{11}+c_{22}+c_{33})+2/9(c_{12}+c_{13}+c_{23})$
**$n$**$= (3B-2G)/[2(3B+G)]$
$E=(9*B*G)/(3*B+G)$

**Table 2: Pressure derivatives of frequencies of phonons of LiYF$_4$ at 0 K at low pressure (below 8 GPa) and high pressure (above 8 GPa).**

| Phonon modes | d?/dP (cm-1/GPa) [0-8] Exp. | d?/dP (cm-1/GPa) [0-8] Calc. | d?/dP (cm-1/GPa) [10-16] Calc. |
|---|---|---|---|
| Ag | - | 0.23 | -0.34 |
| Eg | - | -0.29 | -0.38 |
| Bg | - | 0.33 | -0.50 |
| Eg | - | 2.33 | 1.97 |
| Bg | 1.13 | 1.07 | 0.53 |
| Ag | 5 | 5.07 | 3.75 |
| Eg | 6.6 | 4.47 | 2.67 |
| Bg | 6.6 | 4.86 | 2.00 |
| Bg | 4.54 | 2.33 | 3.92 |
| Eg | 4.54 | 4.57 | 4.30 |
| Bg | 5.77 | 5.53 | 4.12 |
| Ag | 5.77 | 10.37 | 8.93 |
| Eg | 14.43 | 13.53 | 9.69 |

**Figure 1**: Lattice parameters vs. pressure for the I4$_1$/a (Z=4), I2/a (Z=4) and P2$_1$/c (Z=4) initial structures.

**Figure 2**: X-ray powder diffraction patterns of P1 optimized structures at different pressures. The initial structures were built from (a) I4$_1$/a, (b) I2/a and (c) P2$_1$/c space group structures.

**Figure 3**: Interatomic distances vs. pressure for the P2$_1$/c (Z=4) initial structures. (a) Li-F (b) Y-F

**Figure 4**: Unit cell of P2$_1$/c structure optimized at 20 GPa. Little size Black atoms correspond to F anions, little size grey atoms to Li cations and Big size light grey atoms to Y cations. The LiF$_5$ bipyramide is shown in (a) and YF$_8$ octahedra in (b).

**Figure 5**: Angle F vs. pressure. The scheme represents the projection of the tetrahedral in a unit cell of LiYF$_4$. on the (010) plane. The black atoms correspond to F anions.

**Figure 6**: Enthalpy vs. pressure. The enthalpy of $S_P^{I4_1/a}$ are taken as reference.

**Figure 7**: Elastic constants vs. pressure of scheelite LiYF$_4$.

**Figure 8**: Pressure dependence of the Raman frequencies of scheelite LiYF$_4$.

**Figure 9**: Phonon dispersion curves of scheelite LiYF$_4$ for 0 GPa, 4 GPa, 8 GPa, 12 GPa, 16 GPa, 20 GPa. The path is defined in direction of the quadratic Brillouin zone of the scheelite structure. The labels are adapted to the symmetry of the structure. Z (1/2 1/2 –1/2), G=$\Gamma$ (0 0 0), X (0 0 1/2), P(1/4 1/4 1/4), N(0 0.5 0)

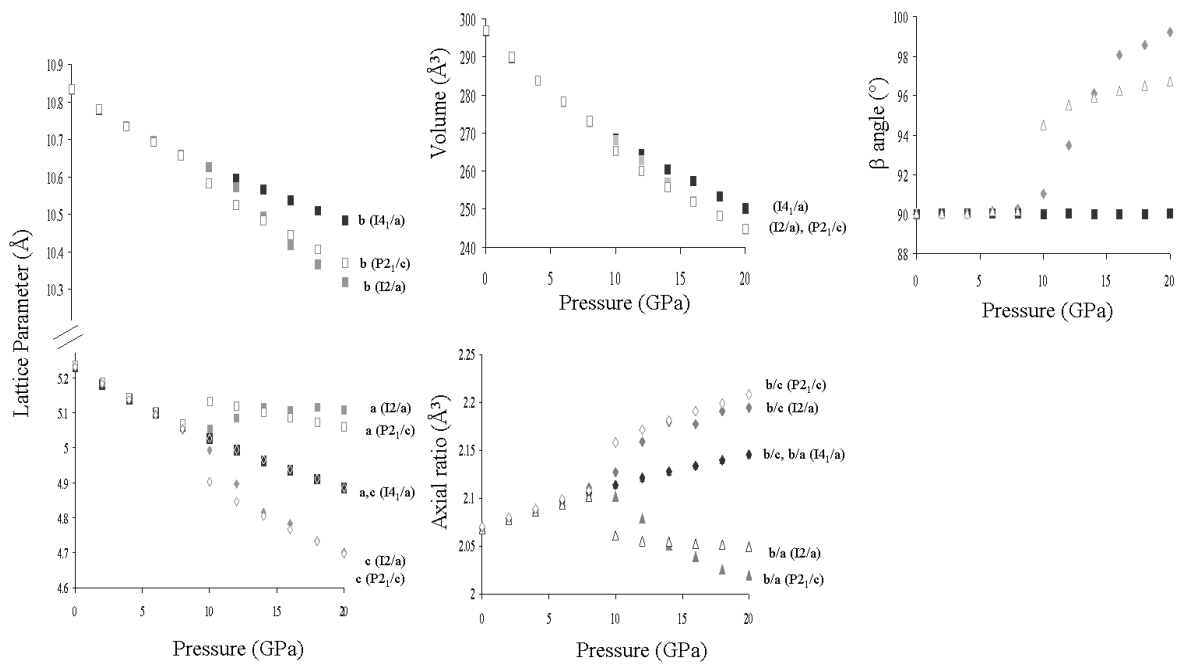

**Figure 1 : Lattice parameters vs. pressure for the $I4_1/a$ (Z=4), $I2/a$ (Z=4) and $P2_1/c$ (Z=4) initial structures.**

**Figure 2 : X-ray powder diffraction patterns of P1 optimized structures at different pressures. The initial structures were built from (a) I4$_1$/a, (b) I2/a and (c) P2$_1$/c space group structures.**

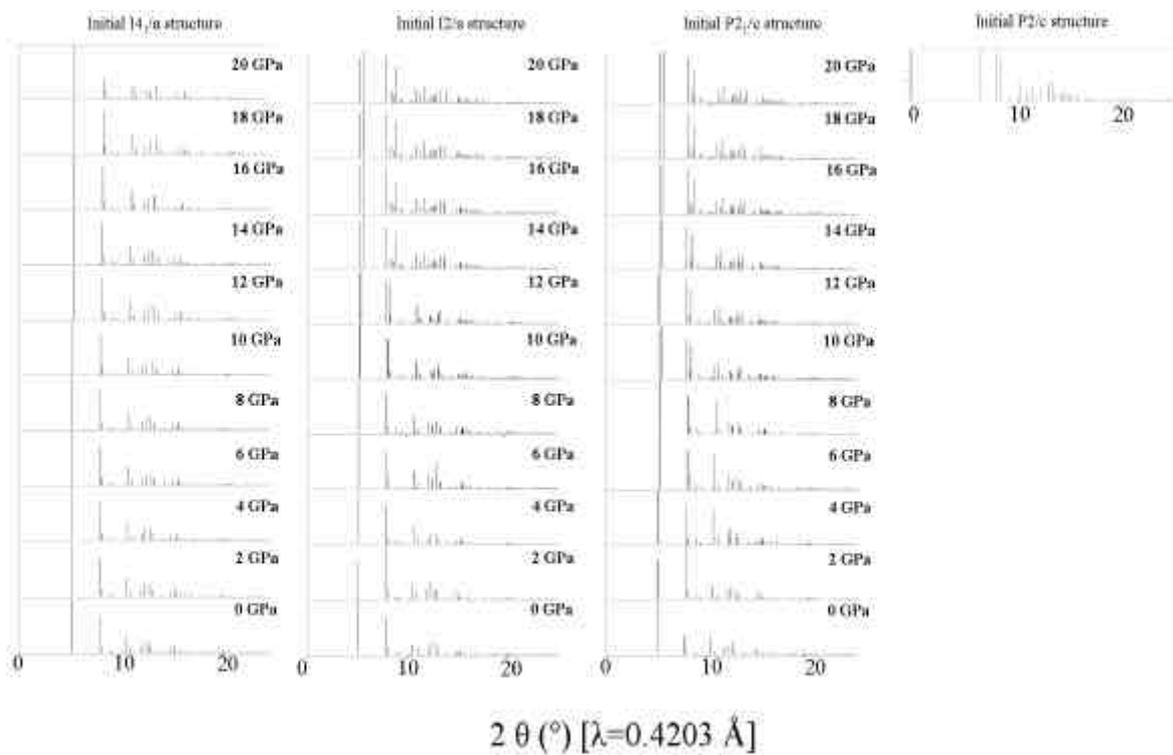

$2\theta\ (°)\ [\lambda=0.4203\ \text{Å}]$

**Figure 3 : Interatomic distances vs. pressure for the P2$_1$/c (Z=4) initial structures. (a) Li-F (b) Y-F**

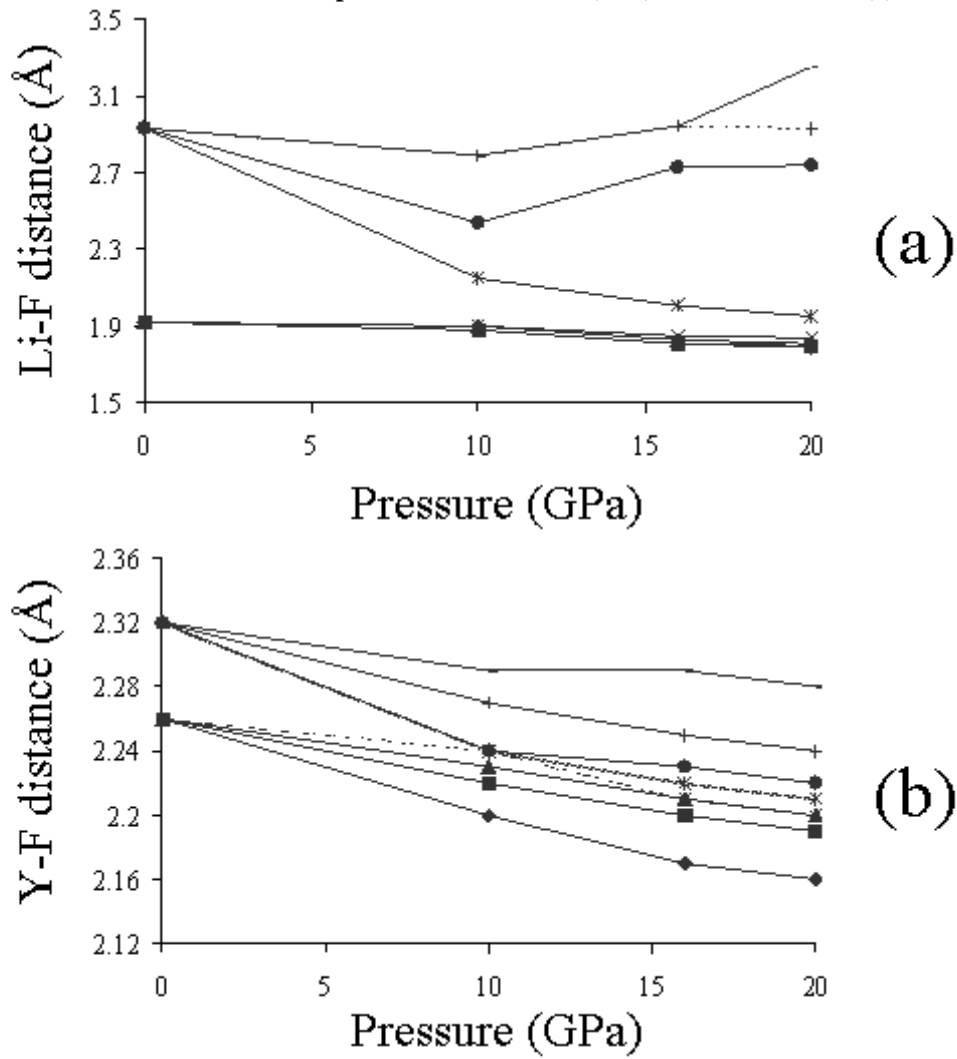

**Figure 4 : Unit cell of P2₁/c structure optimized at 20 GPa. Little size Black atoms correspond to F anions, little size grey atoms to Li cations and Big size light grey atoms to Y cations. The LiF₅ bipyramide is shown in (a) and YF₈ octahedra in (b).**

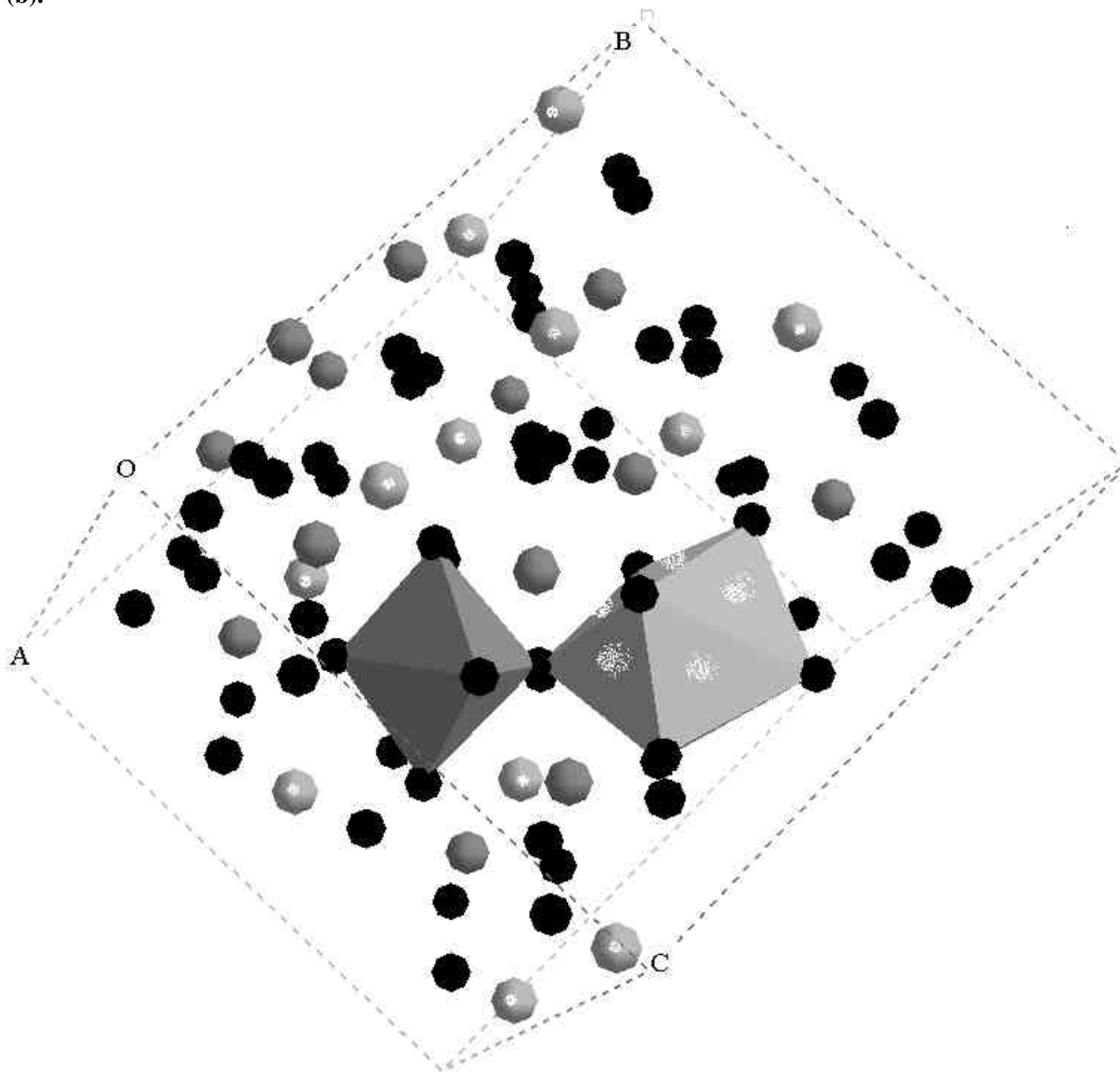

**Figure 5 : Angle F vs. pressure. The scheme represents the projection of the tetrahedral in a unit cell of LiYF$_4$. on the (010) plane. The black atoms correspond to F anions.**

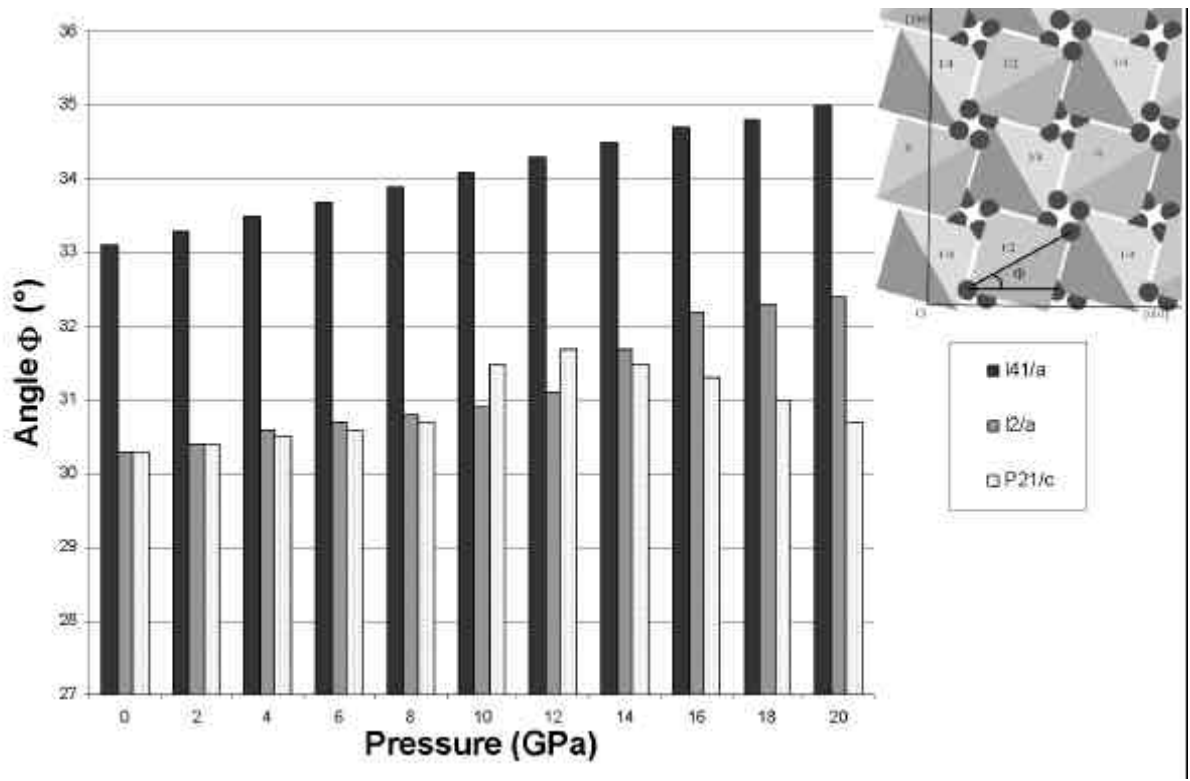

**Figure 6 : Enthalpy vs. pressure. The enthalpy of $S_P^{I4_1/a}$ are taken as reference.**

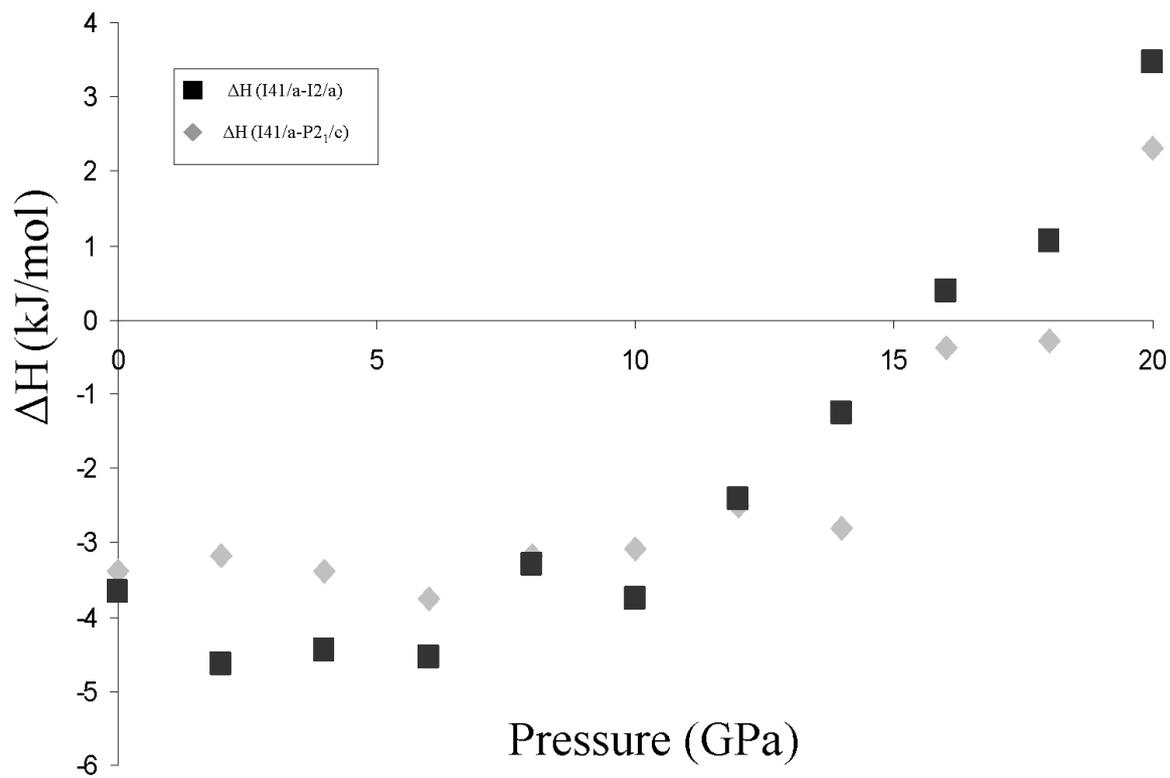

**Figure 7 : Elastic constants vs. pressure of scheelite LiYF$_4$.**

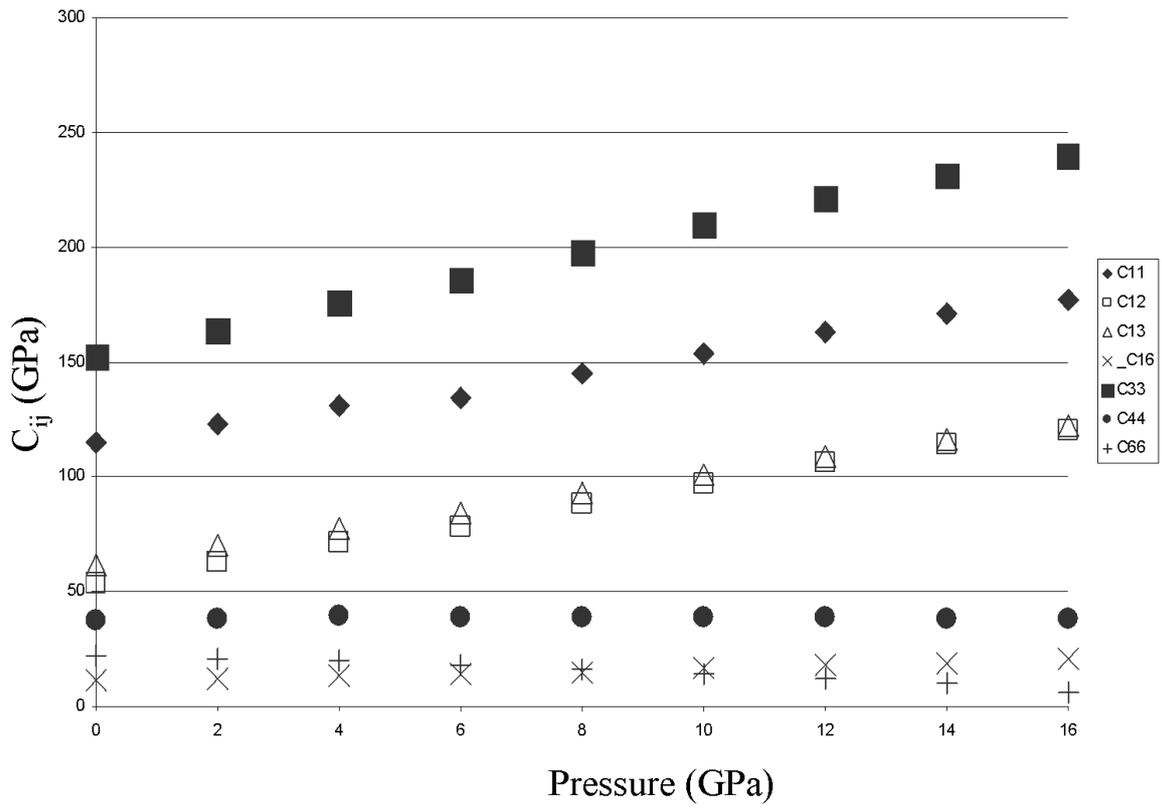

**Figure 8 : Pressure dependence of the Raman frequencies of scheelite LiYF$_4$.**

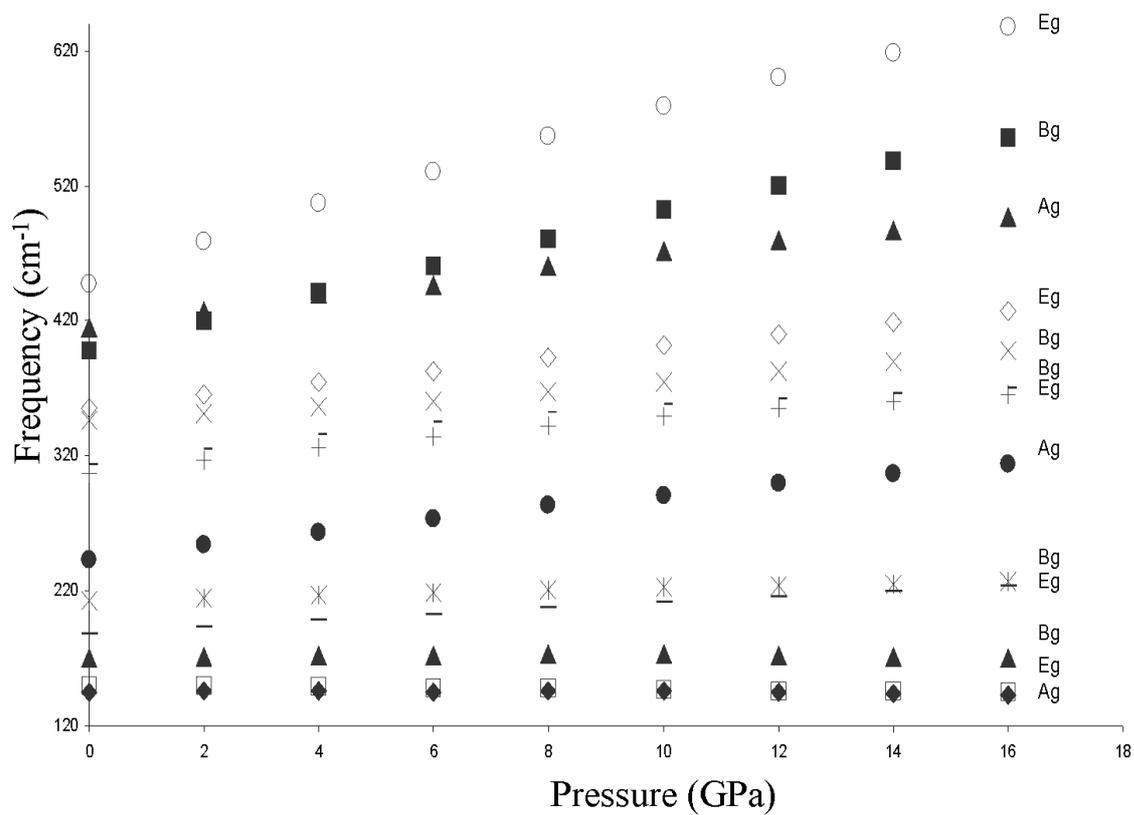

**Figure 9 : Phonon dispersion curves of scheelite LiYF$_4$ for 0 GPa, 4 GPa, 8 GPa, 12 GPa, 16 GPa, 20 GPa. The path is defined in direction of the quadratic Brillouin zone of the scheelite structure. The labels are adapted to the symmetry of the structure. Z (1/2 1/2 –1/2), G=Γ (0 0 0), X (0 0 1/2), P(1/4 1/4 1/4), N(0 0.5 0)**

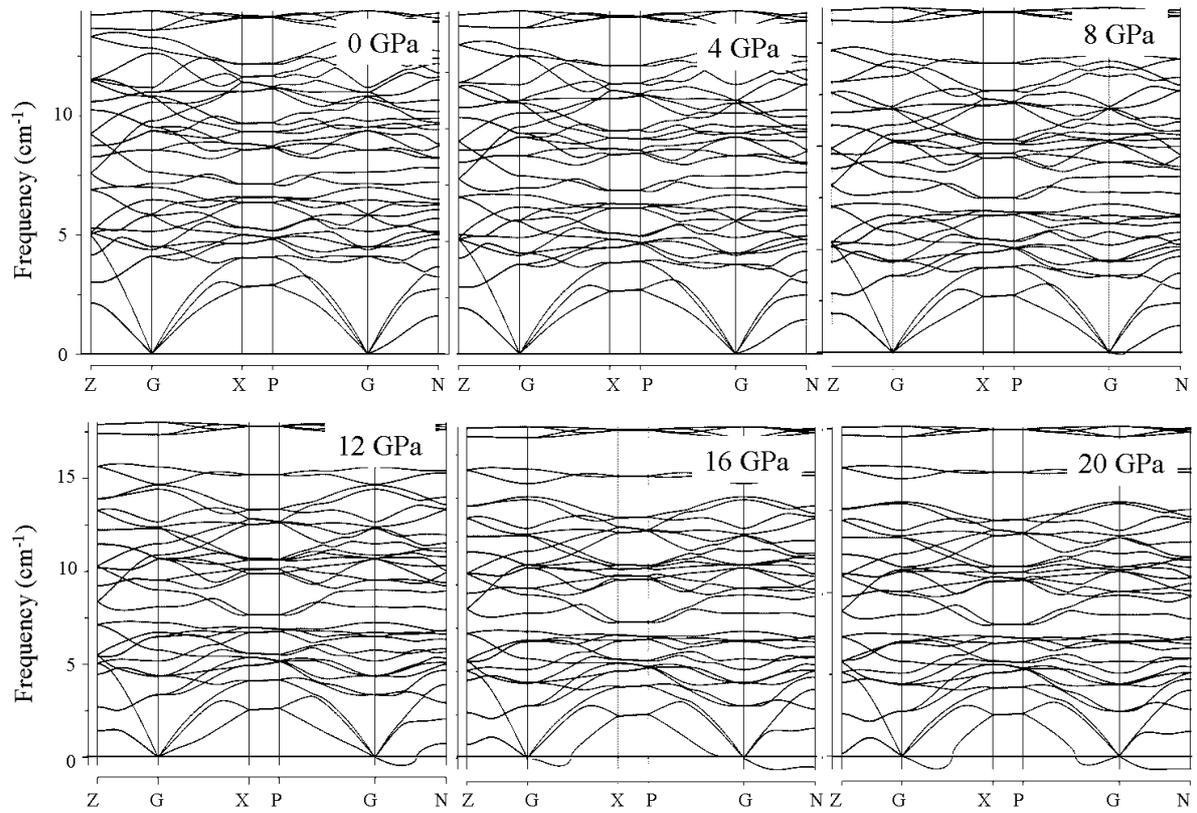